\documentclass[12pt,pra,aps]{revtex4}
\usepackage{amssymb}
\usepackage{amsmath}
\usepackage{amsfonts}
\usepackage{graphicx}
\usepackage{bbold}

\newtheorem{theorem}{Theorem}[section]

\newtheorem{proposition}[theorem]{Proposition}

\newenvironment{proof}[1][Proof]{\begin{trivlist}
\item[\hskip \labelsep {\bfseries #1}]}{\end{trivlist}}
\newenvironment{definition}[1][Definition]{\begin{trivlist}
\item[\hskip \labelsep {\bfseries #1}]}{\end{trivlist}}

\newcommand{\qed}{\nobreak \ifvmode \relax \else
      \ifdim\lastskip<1.5em \hskip-\lastskip
      \hskip1.5em plus0em minus0.5em \fi \nobreak
      \vrule height0.75em width0.5em depth0.25em\fi}

\begin{document}

\title{A category theoretic approach to unital quantum channel convexity and Birkhoff's theorem}

\author{Ian T. Durham}
\email[]{idurham@anselm.edu}
\affiliation{Department of Physics, Saint Anselm College, Manchester, NH 03102}
\date{\today}

\begin{abstract}
Birkhoff's Theorem states that doubly stochastic matrices are convex combinations of permutation matrices.  Quantum mechanically these matrices are doubly stochastic channels, i.e. they are completely positive maps preserving both the trace and the identity.  We expect these channels to be convex combinations of unitary channels and yet it is known that some channels cannot be written that way.  Recent work has suggested that $n$ copies of a single channel might approximate a mixture (convex combination) of unitaries.  In this paper we show that $n(n+1)/2$ copies of a symmetric unital quantum channel may be arbitrarily-well approximated by a mixture (convex combination) of unitarily implemented channels.  In addition,  we prove that any extremal properties of a channel are preserved over $n$ (and thus $n(n+1)/2$) copies.  The result has the potential to be completely generalized to include non-symmetric channels.
\end{abstract}


\maketitle

\section{Introduction and background}
There is a famous theorem attributed to Garrett Birkhoff that states that doubly stochastic matrices are convex combinations of permutation matrices.  In the quantum context, doubly stochastic matrices become doubly stochastic channels, i.e. completely positive maps preserving both the trace and the identity.  Quantum mechanically we understand the permutations to be the unitarily implemented channels.  That is, we expect doubly stochastic quantum channels to be convex combinations of unitary channels.  Unfortunately it is well-known that some quantum channels $cannot$ be written that way \cite{Landau:1993yq,Gregoratti:2003rt}.  Recent work has suggested that $n$ copies of a single channel might approximate a mixture (convex combination) of unitaries \cite{Winter:fr,Smolin:2005mz,Mendl:2008gf,Watrous:2008ul}.  In this article we prove a slightly stronger result for symmetric unital quantum channels via the following theorem.
\begin{theorem}
Given a symmetric unital quantum channel that maps from $\rho$ to $T(\rho)$, we may approximate $n(n+1)/2$ copies of such a channel arbitrarily well with a mixture (convex combination) of unitarily implemented channels.
\end{theorem}
In addition, we demonstrate that extremal properties of unital quantum channels are maintained over $n(n+1)/2$ copies and that such channels behave classically in the limit of classical information.

We prove the aforementioned theorem in several stages.  First we discuss convexity beginning with a review of the classical version of Birkhoff's theorem.  We then discuss convexity as it relates to unital quantum channels and prove that $n$ (and thus $n(n+1)/2$) copies of a unital quantum channel $T$ retain any extremal properties $T$ might possess, where $n$ has no restrictions.  After a brief review of basic category theory, we demonstrate that a single symmetric unital quantum channel along with the associated Hilbert space is a category with one object.  By definition this also means that it is a group.  Cayley's theorem then implies that it is also isomorphic to some permutation group.  We then construct a larger category consisting of $n(n+1)/2$ copies of our unital quantum channel with this category (which consists of groups) being isomorphic to a set of permutation groups.  Finally we apply a group theoretic result to show that these permutation groups are, in fact, unitary.

\section{Convexity}
We begin by reviewing Birkhoff's theorem (also known as the Birkhoff-von Neumann theorem).  It may be stated in a number of ways and our presentation follows closely that given in Steele\cite{Steele:2004rz}.  Given a permutation $\sigma \in S_n$, the permutation matrix that is associated with $\sigma$ is the $n \times n$ matrix $P_{\sigma}(j,k)$, where $1 \le j,k \le n$, whose entries are given by
\begin{equation*}
P_{\sigma}(j,k)=
\left \{
\begin{array}{c c}
1 & \textrm{if } \sigma (j)=k \\
0 & \textrm{otherwise}.
\end{array}
\right.
\end{equation*}
An $n \times n$ doubly stochastic matrix $D$ is a square matrix whose elements are real and whose rows and columns sum to unity.  Given such a matrix, there exist nonnegative weights $\{w_{\sigma}:\sigma \in S_n\}$ such that
\begin{equation}
\begin{array}{c c c}
\sum_{\sigma \in S_n} w_{\sigma}=1 & \quad \textrm{and} \quad & \sum_{\sigma \in S_n} w_{\sigma} P_{\sigma}=D.
\end{array}
\end{equation}
In other words, every doubly stochastic matrix is an average (convex combination of) permutation matrices.  The set of such matrices of order $n$ is said to form the convex hull of permutation matrices of the same order where the latter are the vertices (extreme points) of the former.  It superficially appears as if this ought to translate quite easily into the quantum world, particularly if we were to assume that the nonnegative weights were actually probabilities.  Unfortunately that is not the case.

In the quantum world we are particularly interested in what a matrix like $D$ can $do$.  In other words, if we treat it like a map or a `black box'\footnote{Pedagogically we often make use of an old-fashioned meat grinder as a visual cue, inspired in part by an old $Monty$ $Python$ animated skit.  We call it our `quantum meat grinder.'} we wish to note what sort of an output we get given a specific input.  Generally we are interested in completely positive (CP) maps which are maps preserving both trace and identity (i.e. and identity preserving CPTP map).  They are represented by square matrices of dimension $d$ and are known as unital quantum channels.  They can be written as a \emph{Kraus decomposition} as
\begin{eqnarray}
T(\rho)=\sum_{i}A_{i}\rho A_{i}^{\dag} &
\qquad \textrm{where} \qquad &
\sum_{i} A_{i}^{\dag}A_{i} = {\mathbb 1}
\end{eqnarray}
and $T(\mathbb{1})={\mathbb 1}$.  The set of all quantum channels on $\mathcal{M}_{d}$ is convex and compact meaning it may be decomposed as
\begin{equation}
T=\sum_{i}p_{i}T_{i}
\end{equation}
where the $p$'s are probabilities and the $T_{i}$'s are $extremal$ unital channels, that is channels that may not be further decomposed \cite{Mendl:2008gf}.  Channels with a single Kraus operator are pure channels and the extremal points in the convex set of channels are precisely the pure channels.  Here $T$ represents the set of $all$ channels on the particular space, not necessarily copies of the same one, i.e. the $T_i$ may not represent the same channel.  $T$ also has additional restrictions not possessed by $D$.  

\subsection{Preservation of extremal properties for unital quantum channels}
The preservation of unitality over tensor products is a basic property of the tensor product of algebras.  As such this implies that the trace-preserving property is itself preserved, i.e. 
\begin{equation}
\sum_{i_1,\ldots,i_n} (A_{i_1}\otimes\cdots\otimes 
A_{i_n})^\dagger(A_{i_1}\otimes\cdots\otimes A_{i_n})= \textrm{id}
\end{equation}
However, it is not clear that the extremal properties, if possessed, are also preserved.  Specifically, Landau and Streater have proven a theorem that if $T$ is unital, it is extremal if and only if the set
\begin{equation}
\left \{A_{k}^{\dag}A_{l} \oplus A_{l}A_{k}^{\dag} \right \}_{k,l \ldots N}
\end{equation}
is linearly independent \cite{Landau:1993yq}.  Thus, as we take a tensor product of $T$ with itself, if we wish for the resultant channel to remain extremal we require that the set
\begin{equation}
\{(A_{k_1}^\dagger\otimes A_{k_2}^\dagger)(A_{l_1}\otimes A_{l_2})
\oplus
(A_{l_1}\otimes A_{l_2})(A_{k_1}^\dagger\otimes 
A_{k_2}^\dagger)\}_{k_1,k_2,l_1,l_2}
\end{equation}
remain linearly independent.  Generalized over $n$ tensor products we require that
\begin{equation}
\left \{\left(A_{k_{1}}^{\dag}\otimes\cdots\otimes A_{k_{n}}^{\dag}\right)\left(A_{l_{1}}\otimes\cdots\otimes A_{l_{n}}\right)\oplus \left(A_{l_{1}}\otimes\cdots\otimes A_{l_{n}}\right)\left(A_{k_{1}}^{\dag}\otimes\cdots\otimes A_{k_{n}}^{\dag}\right) \right \}_{k,l \ldots N}
\end{equation}
be linearly independent.  This is amounts to showing that equation (7) is of the same basic form as equation (5).  In other words, we require that
\begin{equation}
\begin{array}{ccl}
\left(A_{k_{1}}^{\dag}\otimes\cdots\otimes A_{k_{n}}^{\dag}\right) & \left(A_{l_{1}}\otimes\cdots\otimes A_{l_{n}}\right) & =\left(A_{j}^{\dag}A_{m}\right)_{j,m\ldots N} \\
\left(A_{l_{1}}\otimes\cdots\otimes A_{l_{n}}\right) & \left(A_{k_{1}}^{\dag}\otimes\cdots\otimes A_{k_{n}}^{\dag}\right) & =\left(A_{m}A_{j}^{\dag}\right)_{j,m\ldots N}.
\end{array}
\end{equation}
In the case that the Kraus operators are symmetric this can be easily accomplished via a Cholesky decomposition in which any symmetric, square, positive definite matrix can be decomposed into the product of a lower triangular matrix and its conjugate transpose (there is also a version that uses an upper triangular matrix) while preserving linear independence \cite{Watkins:2002zl}.  In other words, as long as the matrices on the left side of equation (8) are symmetric, Cholesky decomposition may be used to find equations of the form required by the right side of equation (8).  Even though Kraus operators are not necessarily unique for a given quantum operation, all systems of Kraus operators that represent the same quantum operation are related via a unitary transformation (see Theorem 8.2 in \cite{Nielsen:2000rz}).  

Suppose we have a set of Kraus operators for a given unital quantum channel that is known to be extremal.  Since extremality is a unique metric property there will be a unique value for the fidelity of the channel if it is extremal.  Since fidelity is invariant under unitary transformations and the fidelity associated with an extremal channel is unique, extremality is preserved under such transformations.  Thus, even though the Kraus operators may not be unique for a given extremal channel, the Cholesky decomposition as applied in equation (8) will preserve extremality for the set of all extremal Kraus operators for that channel.

Note that while Cholesky decomposition only works for symmetric matrices, this doesn't severely limit the relevance of these results since there do exist symmetric channels that violate Birkhoff's theorem.  An example of one such channel is the `A not Q' channel given in \cite{Gregoratti:2003rt}.  Since this channel is diagonal it is symmetric and thus Cholesky decomposition may be used to prove (8) and, by the above argument regarding fidelity, extremality is preserved. 

In general, we expect that most quantum channels are symmetric if they are to preserve identity, though this is not a foregone conclusion (i.e. there are matrices that could satisfy equation (2), preserving identity, that are not symmetric, though we doubt any such matrices are physically significance).  At any rate, it is presently unknown (at least to this author) whether a method exists to similarly decompose \emph{non-symmetric} (or non-diagonalizable) Kraus operators for a quantum channel while retaining the extremal properties in the asymptotic limit, i.e. replicating equation (8).

In any case, we have at least shown that \emph{symmetric} channels, including certain troublesome channels (e.g. the `A not Q' channel mentioned above), retain their extremal characteristics in the asymptotic limit.  This is an important step in verifying the suggested quantum extension of Birkhoff's theorem.

\section{A review of categories}

We now pause for a brief review of category theory.  (The reader is encouraged to consult a standard text on the subject \cite{Maclane:1971lq,Awodey:2006rr} as well as materials introducing it to quantum theory \cite{Coecke:2006wd,Abramsky:2004eu,Abramsky:2007qe,Baez:2006ai}.)  The following definitions closely follow those given in Awodey\cite{Awodey:2006rr}.
\begin{definition}[Category]
A $category$ is a mathematical structure that consists of $Objects$: $A,B,C,\ldots$ and $Arrows$: $f,g,h,\ldots$ with the properties,
\begin{itemize}
\item
for each arrow $f$ there are given objects dom($f$) and cod($f$) called the $domain$ and $codomain$ of $f$ respectively.  We write $f: A \to B$ to indicate that $A=$ dom($f$) and $B=$ cod($f$);
\item
given arrows $f: A \to B$ and $g: B \to C$ where cod($f$) = dom($g$) there is given an arrow $g \circ f: A \to C$ called the $composite$ of $f$ and $g$;
\item
for each object $A$ there is given an arrow ${\mathbb 1}_{A}: A \to A$ called the $identity$ $arrow$ of $A$.
\end{itemize}
These must satisfy
\begin{itemize}
\item
\textbf{Associativity} $h \circ (g \circ f) = (h \circ g) \circ f$ for all $f: A \to B, g: B \to C, h: C \to D$;
\item
\textbf{Unit} $f \circ {\mathbb 1}_{A}=f={\mathbb 1}_{B} \circ f$ for all $f: A \to B$.
\end{itemize}
\end{definition}
\begin{definition}[Functor]
A $functor$ $F:\textrm{\textbf{C}} \to \textrm{\textbf{D}}$ between categories \textbf{C} and \textbf{D} is a mapping from objects to objects and arrows to arrows such that
\begin{itemize}
\item
$F(f: A \to B)=F(f): F(A) \to F(B)$,
\item
$F(g \circ f)=F(g) \circ F(f)$,
\item
$F({\mathbb 1}_{A})={\mathbb 1}_{F(A)}$.
\end{itemize}
\end{definition}
\begin{definition}  In any category $\textrm{\textbf{C}}$, an arrow $f: A \to B$ is called an isomorphism if there is an arrow $g: B \to A$ in $\textrm{\textbf{C}}$ such that $g \circ f = {\mathbb 1}_{A}$ and $f \circ g = {\mathbb 1}_{B}$.
\end{definition}

In category theory, tensor products form a bifunctor from the category of vector spaces to itself and are covariant in each argument.  As such they may be decomposed as
\begin{equation}
T(\rho) \otimes T(\rho) = (T \otimes T)(\rho \otimes \rho).
\end{equation}
That is, if $T$ is a linear map $T: R \to R$ with $\rho \in R$, then $T \otimes T$ is a linear map $T \otimes T: R \otimes R \to R \otimes R$ and $\rho \otimes \rho \in R \otimes R$.
This is easily generalized to $n$ copies.  In other words, $\rho^{\otimes n}$ is the input state to $n$ channels.  The definition of functors requires that they preserve identity morphisms \cite{Borceaux:1994dp,Maclane:1971lq}.  It is clear that a unital channel is an identity morphism since $T({\mathbb 1})={\mathbb 1}$ and thus $n$ copies of such a channel should preserve this property.  In other words, $T^{\otimes n}({\mathbb 1}^{\otimes n})={\mathbb 1}^{\otimes n}$.

\section{Category and group structure of quantum channels}
A unital quantum channel is a mapping between Hilbert spaces, $\Phi : L(\mathcal{H}_{A}) \to L(\mathcal{H}_{B})$, where $L(\mathcal{H}_{i})$ is the family of operators on $\mathcal{H}_{i}$.  The operator spaces can be interpreted as $C^{*}$-algebras and thus we can also view the channel as a mapping between $C^{*}$-algebras, $\Phi : \mathcal{A} \to \mathcal{B}$.  Quantum channels can carry classical information as well.  An example of such a channel would be $\Phi : L(\mathcal{H}_{A}) \otimes C(X) \to L(\mathcal{H}_{B})$ where $C(X)$ is the space of continuous functions on some set $X$ and is also a $C^{*}$-algebra.  In other words, whether or not classical information is processed by the channel, it (the channel) is a mapping between $C^{*}$-algebras.  Note, however, that these are not necessarily the same $C^{*}$-algebras.  Since the channels are represented by square matrices, the input and output $C^{*}$-algebras must have the same dimension, $d$.  Thus we can consider them both subsets of some $d$-dimensional $C^{*}$-algebra, $\mathcal{C}$, i.e. $\mathcal{A} \subset \mathcal{C}$ and $\mathcal{B} \subset \mathcal{C}$.  Thus a unital quantum channel is a mapping from $\mathcal{C}$ to itself (the need for a single object will become apparent in a moment).  As such we propose the following:
\begin{proposition}
A unital quantum channel given by $t: L(\mathcal{H}_{\rho}) \to L(\mathcal{H}_{T(\rho)})$, together with the d-dimensional $C^{*}$-algebra, $\mathcal{C}$, on which it acts, forms a category we call $\mathrm{\mathbf{Chan}}(d)$.
\end{proposition}
\begin{proof}
Our object in this case is $\mathcal{C}$ (the $C^{*}$-algebra) while our arrow is $t$ (the channel).  The existence of a domain and codomain of $t$ are trivially obvious.  We must show compositeness.  Consider the unital quantum channels
\begin{eqnarray*}
r: L(\mathcal{H}_{\rho}) \to L(\mathcal{H}_{\sigma}) &
\qquad \textrm{where} \qquad &
\sigma=\sum_{i}A_{i}\rho A_{i}^{\dag} \\
t: L(\mathcal{H}_{\sigma}) \to L(\mathcal{H}_{\tau}) &
\qquad \textrm{where} \qquad &
\tau=\sum_{j}B_{j}\sigma B_{j}^{\dag} 
\end{eqnarray*}
where the usual properties of such channels are assumed (e.g. trace preserving, etc.).  We form the composite $t \circ r: L(\mathcal{H}_{\rho}) \to L(\mathcal{H}_{\tau})$ where
\begin{align}
\tau & = \sum_{j}B_{j}\left(\sum_{i}A_{i}\rho A_{i}^{\dag}\right)B_{j}^{\dag} \notag \\
& = \sum_{i,j}B_{j}A_{i}\rho A_{i}^{\dag}B_{j}^{\dag} \\
& = \sum_{k}C_{k}\rho C_{k}^{\dag} \notag
\end{align}.  Since $A$ and $B$ are summed over separate indices the trace-preserving property is maintained, i.e. $\sum_{k} C_{k}^{\dag}C_{k}={\mathbb 1}$  For a similar methodology, see \cite{Nayak:2006rz}.  

There clearly exists an identity arrow in at least one instance due to the nature of unital quantum channels, i.e. by definition $T({\mathbb 1})={\mathbb 1}$.  We may take the completely $general$ identity arrow
\begin{equation*}
{\mathbb 1}_{\rho}: L(\mathcal{H}_{\rho}) \to L(\mathcal{H}_{\rho})
\end{equation*}
to be the time evolution of the state $\rho$ in the $absence$ of any unital quantum channel.  Since this definition is suitably general we have that
\begin{equation*}
t \circ {\mathbb 1}_{A}=t={\mathbb 1}_{B} \circ t \quad \forall \,\, t: A \to B
\end{equation*}
and we are thus left to prove the associativity of composition.

Consider the three unital quantum channels $r: L(\mathcal{H}_{\rho}) \to L(\mathcal{H}_{\sigma})$, $t: L(\mathcal{H}_{\sigma}) \to L(\mathcal{H}_{\tau})$, and $v: L(\mathcal{H}_{\tau}) \to L(\mathcal{H}_{\upsilon})$ where $\sigma=\sum_{i}A_{i}\rho A_{i}^{\dag}$, $\tau=\sum_{j}B_{j}\sigma B_{j}^{\dag}$, and $\upsilon=\sum_{k}C_{k}\tau C_{k}^{\dag}$.  We have
\begin{align}
v \circ (t \circ r) & = v \circ \left(\sum_{i,j}B_{j}A_{i}\rho A_{i}^{\dag}B_{j}^{\dag}\right) = \sum_{k}C_{k} \left(\sum_{i,j}B_{j}A_{i}\rho A_{i}^{\dag}B_{j}^{\dag}\right) C_{k}^{\dag} \notag \\
& = \sum_{i,j,k}C_{k}B_{j}A_{i}\rho A_{i}^{\dag}B_{j}^{\dag}C_{k}^{\dag} = \sum_{i,j,k}C_{k}B_{j}\left(A_{i}\rho A_{i}^{\dag}\right)B_{j}^{\dag}C_{k}^{\dag} \notag \\
& = \left(\sum_{i,j,k}C_{k}B_{j}\tau B_{j}^{\dag}C_{k}^{\dag}\right) \circ r = (v \circ t) \circ r \notag
\end{align}
and thus we have associativity.  $\Box$
\end{proof}

We have intentionally defined our category such that we have a single object, $\mathcal{C}$, rather than two objects, $\rho$ and $T(\rho)$.  A category with one object is known as a \emph{monoid}.  Restricting ourselves to monoids allows us to make use of the following definition.
\begin{definition}
A group $G$ is a category with one object, in which every arrow is an isomorphism \cite{Awodey:2006rr}.  Conversely, any category with a single object whose arrows are invertible is a group under composition \cite{MacLane:1999sp}.
\end{definition}
Note that the invertibility of the arrows in category theoretic terms does not necessarily mean that the channel itself is invertible, i.e. unitary.  If it were, these steps would be unnecessary since there are much simpler ways to show that unitary channels stay unitary in the asymptotic limit.  The invertibility here, however, is at the category-theoretic level as is the isomorphism.  In fact, as Awodey points out, in many cases the \emph{only} definition that makes sense is the abstract, category-theoretic one.  Thus, while the unital quantum channel itself may not be invertible, a category formed with it might be.  Physically, an example of this would be the following.  Suppose we have a non-polarized light beam that passes through a polarizer.  This could be modeled as our quantum channel and is clearly not invertible, i.e. we can't pass the polarized light back through the polarizer to depolarize it.  However, we \emph{could} pass the polarized light through an optical depolarizer.  In a category theoretic sense we would still only have one object, though we would now have two arrows that satisfy the above definition of a category-theoretic isomorphism.  As such, our category \textbf{Chan}($d$), is a group.  This allows us to make use of a powerful theorem of group theory known as Cayley's theorem.  
\begin{theorem}[Cayley] Every group is isomorphic to a permutation group.
\end{theorem}
There are numerous presentations and proofs of this theorem.  Since it is well-known, we refer the reader to one of the following for a proof \cite{Wallace:1998th,Beachy:2006fy,Armstrong:1997la,Awodey:2006rr}.

This means that a unital quantum channel and the associated $C^{*}$-algebra \emph{is isomorphic to a permutation group}.  We might think we can simply say now that $n$ copies of this channel would be isomorphic to a set of permutation groups and leave it at that.  The problem is that the dimension of the $C^{*}$-algebra is dependent on the value of $n$ because two copies of a channel are connected by a tensor product.

Let us then define \textbf{Chan}($d$) to be a category with a $d$-dimensional $C^{*}$-algebra as the sole object.  So, for example, suppose our basic channel is associated with a two-dimensional $C^{*}$-algebra.  It's category would be \textbf{Chan}(2).  Two copies of this channel exist are associated with a four-dimensional $C^{*}$-algebra and so their category would be \textbf{Chan}(4).  We can generalize this a bit more by defining \textbf{Chan}($d^{n}$) to be a category with a $d^n$-dimensional $C^{*}$-algebra.

Categories are like groups in that one can have a category made up of categories.  Thus we define the category \textbf{Chan} with objects $\textrm{\textbf{Chan}}(d),\textrm{\textbf{T}}(d^{2}),\ldots,\textrm{\textbf{Chan}}(d^{n})$ and the arrow $\otimes$.  Since tensor products are known to be associative, we can move between the objects in our new category fairly easily.  For example
\begin{equation}
[\textrm{\textbf{Chan}}(d) \overset{\otimes}{\longrightarrow} \textrm{\textbf{Chan}}(d^{2})] \overset{\otimes}{\longrightarrow} \textrm{\textbf{Chan}}(d^{3})=\textrm{\textbf{Chan}}(d) \overset{\otimes}{\longrightarrow} [\textrm{\textbf{Chan}}(d^{2}) \overset{\otimes}{\longrightarrow} \textrm{\textbf{Chan}}(d^{3})].
\end{equation}
Note in this case that the arrow is not an isomorphism.  This is fine since the isomorphism criterion was only needed in order to utilize Cayley's theorem to ensure that each \textbf{Chan}($d^{n}$) was isomorphic to a permutation group.  Also notice that the category \textbf{Chan} contains $n(n+1)/2$ copies of our original channel rather than simply $n$.  We will discuss the importance of this after proving the main theorem of this article.

We are now in a position to prove the central theorem of our paper.
\begin{theorem}
Given a symmetric unital quantum channel that maps from $\rho$ to $T(\rho)$, we may approximate $n(n+1)/2$ copies of such a channel arbitrarily well with a mixture (convex combination) of unitarily implemented channels.
\end{theorem}
\begin{proof}
By Cayley's theorem we've shown that each category, \textbf{Chan}($d^{n}$), which is also a group, is individually isomorphic to some permutation group.  As a group, \textbf{Chan}($d^{n}$) has infinite cardinality since we're invoking the $C^{*}$-algebra for the entire space of inputs and outputs.  Thus, since Cayley's theorem assumes an exact isomorphism, the permutation group must also have infinite cardinality.  However, each of these infinite permutation groups is $\varepsilon$-isomorphic to some \emph{finite} permutation group.  Note that any representation of a permutation group that is finite and compact may be considered unitary where a representation is a set of operators representing the permutations, i.e. the operators that produce the permutations \cite{Bohm:1993rz}.  Thus \emph{representations} of finite and compact permutation groups are necessarily unitary.   So, individually, the channels are $\varepsilon$-isomorphic to something that has a unitary representation.  As $n \to \infty$, the cardinality of the \emph{finite} permutation groups increases and thus the isomorphism approaches exact.  The approach is asymptotic as long as $n$ remains finite which it must in order for the representation to be unitary.  Since the category \textbf{Chan}, while containing $n$ copies of \textbf{Chan}($d^{n}$), actually contains $n(n+1)/2$ copies our unital quantum channel, it is $n(n+1)/2$ copies of that channel that approaches a representation of a set of unitarily represented groups approximately well.  Regarding the term \emph{arbitrarily well}, we interpret this to mean that we may choose \emph{any} value of $\varepsilon$, no matter how small, and still find a better approximation.  This fact remains fundamentally true since $n$ can increase indefinitely while remaining finite. $\Box$
\end{proof}

We note that there are different levels of isomorphism present throughout this proof.  Some are on a category theoretic-level while others are at a group-theoretic level.  This is an important distinction that can affect the interpretation of these results.  As an analogy, a Euro and a Dollar are related by dint of the fact that both are the basis of a particular currency.  Conversely, within, say, the American system, one could possess either a paper dollar or a dollar coin.  This is an entirely different level of relation.

\section{Discussion}
These results represent the most general version of Birkhoff's theorem in the asymptotic limit yet discovered, though we readily admit that the level of abstraction used (including the representation theoretic aspects in the proof of the final theorem) may leave doubts in some readers' minds.  On the other hand, when considering the physical extension of these abstractions, we believe the result captures the essence of the suggested asymptotic limit to Birkhoff's theorem.  In particular, considering the fact that we have an isomorphism to a \emph{representation} in the final proof, we note that, physically, Birkhoff's theorem only implies that we may, in essence, \emph{represent} $n$ copies of a unital quantum channel with a convex combination of unitarily implemented ones.  In other words, we could build something in a lab that approximates $n$ copies of some, perhaps complicated, unital quantum channel.  So we interpret `to approximate' as meaning essentially the same thing as `to represent.'  This is, of course, simply an interpretation and we believe that it would greatly benefit from experimental results.  In addition, as we have noted, there are different levels of isomorphism being used here and it is important to note the differences when interpreting the results.

The results are of particular importance for physical realizations of certain `problematic' unital quantum channels, i.e. those that can't be approximated by unitary channels individually.  It implies that we can approximate $n$ copies of them unitarily allowing us to study their behavior and approximate that behavior in realistic quantum information processing systems.  In particular, a quantum channel allows for perfect environment-assisted error correction if and only if it is a mixture of unitaries.  As such, this result implies we can approximate a system with perfect environment-assisted error correction with a set of unitarily implemented channels for which many relatively simple physical manifestations already exist.

In addition we note that this result is stronger than the original suggestion that the best we could approximate was $n$ copies of a given channel.  Our result implies we may actually approximate $n(n+1)/2$ copies.  While technically larger than $n$ in the asymptotic limit, the number of computational steps required to reach a very large number of copies of our channel is \emph{less}.  This is because each individual category is represented by a single unitarily implemented channel and thus $n$ such channels approximate $n(n+1)/2$ unital quantum channels.

Finally, we note that the restriction to symmetric channels is in order to preserve any extremal characteristics.  If a method is found that could decompose a non-symmetric matrix into a form similar to the right-hand side of equation (8), then our result would be completely general.

\begin{acknowledgements}
We thank Aram Harrow and Steve Shea for in depth and on-going discussions of this problem.  Additionally, we thank Michael Ben-Or and three anonymous reviewers who provided very useful comments.  We also acknowledge several informal conversations concerning various aspects of this paper with Dave Bacon, Winton Brown, Lorenza Viola, and attendees of one of the physics/mathematics seminars at Saint Anselm College.
\end{acknowledgements}

\bibliographystyle{apsrev}
\bibliography{CatBirkJMP.bbl}
\end{document}